\documentstyle[preprint,aps]{revtex}

\def \lapx {{\hbox {$\  {<\atop \sim}\ $}}}

\def \ad {\alpha_d}
\def \al {\alpha_{\ell}}
\def \dal {(\alpha_{\ell}-1.0)}
\def \dad {(\alpha_d-0.7)}
\def \vopo {(VO)$_2$P$_2$O$_7$ \ }
\def \vopon {(VO)$_2$P$_2$O$_7$ \ }
\def \vopoh {VO(HPO$_4$) $\cdot \frac{1}{2} $ H$_2$O \ }
\def \vopohn {VO(HPO$_4$)$\cdot \frac{1}{2} $ H$_2$O \ }

\begin{document}
\draft

\preprint{ORNL-CCIP-94-04 / RAL-94-027}
\title{
The susceptibility and excitation spectrum of \vopo
in ladder and dimer chain models
}

\author{T.Barnes$^{1,2}$ and J.Riera$^{\ast1,3}$}

\address{
$^1$Physics Division and Center for Computationally Intensive Physics,
Oak Ridge National Laboratory, Oak Ridge, TN 37831   \\
\noindent
$^2$Department of Physics and Astronomy,
University of Tennessee, Knoxville, TN 37996      \\
\noindent
$^3$Department of Physics and Astronomy,
Vanderbilt University, Nashville, TN 37235
}

\date{April 1994}

\maketitle

\begin{abstract}
We present numerical results for the magnetic susceptibility of a Heisenberg
antiferromagnetic spin ladder, as a function of temperature and the
spin-spin interaction strengths $J_\perp$ and $J_{||}$.
These are contrasted with new bulk limit results for the dimer chain.
A fit to the
experimental susceptibility of the candidate spin-ladder compound
vanadyl pyrophosphate,
(VO)$_2$P$_2$O$_7$, gives the parameters
$J_\perp = 7.82$ meV and
$J_{||} = 7.76$ meV. With these values we predict
a singlet-triplet energy gap of
$E_{gap} = 3.9$ meV, and give a numerical
estimate of the ladder triplet dispersion
relation $\omega(k)$. In contrast, a fit to the
dimer chain model leads to $J_1=11.11$ meV
and $J_2=8.02$ meV, which predicts a gap of
$E_{gap} = 4.9$ meV.
\end{abstract}

\noindent
$\ast$) Permanent address:
Departamento de F\'isica, Facultad de Ciencias Exactas,  \\
Av. Pellegrini 250, 2000 Rosario, Argentina

\pacs{PACS Indices: 74.65+n, 75.10Jm, 75.30Ds, 75.50Ee. \hfill }

\centerline{\bf I.Introduction}
\vskip 0.25cm

Quantum antiferromagnetism in lower-dimensional systems
has proven to be a very rich subject. Some of the more
dramatic developments include the
realization
that half-integral and integral spin chains have very different
excitation spectra,\cite{Haldane} and evidence that
two-dimensional antiferromagnetism is a crucial component of high temperature
superconductivity.\cite{revtb,reved,revem}

The Heisenberg spin ladder
is interesting theoretically as an intermediary
between half-integer $(S=1/2)$ and integer
$(S=0,1)$ spin chains.
This system has isotropic
nearest-neighbor interactions along the chains ($J_{||}$)
and along the rungs ($J_\perp $) of a
ladder geometry,
\begin{eqnarray}
H =
 J_{||}\sum_{\leftrightarrow} {\bf S}_i \cdot {\bf S}_j
+ J_\perp \sum_{\updownarrow} {\bf S}_i \cdot {\bf S}_j \ .
\end{eqnarray}

Previous studies of the
Heisenberg spin ladder have discussed the ground state energy and the
dependence of the singlet-triplet energy gap
on $J_\perp / J_{||} \equiv \alpha_\ell $
in the antiferromagnetic regime,\cite{DRS,BDRS}
the triplet spin-wave dispersion relation,\cite{BDRS}
the behavior of the system
under doping\cite{DRS} and the dynamical structure function.\cite{BDRS}
Ferromagnetic rung couplings ($J_\perp< 0$, $J_{||}>0$)
have been studied using
Lanczos and Monte Carlo techniques and
the renormalization group;\cite{frung} these references suggest
that a gap exists for all $J_\perp < 0$.

In Ref.\cite{BDRS} we presented numerical and analytical results for the ground
state energy and triplet spin-wave dispersion relation of an $S=1/2$ spin
ladder, as well as numerical results for the structure function $S(\vec
k,\omega)$. We found evidence that a singlet-triplet energy gap appears for any
interchain coupling $J_\perp / J_{||} >0$, and that the spin-wave band minimum
is at $k=\pi$, but the band is folded so the maximum energy occurs between
$k=\pi/2$ (for $J_\perp/J_{||} =0$) and $k=0$ (for $J_\perp/J_{||} =\infty$).
The bandwidth was found to be relatively insensitive to the rung coupling
$J_\perp$, and varied between $\pi J_{||}/2 $ (for $J_\perp/J_{||} =0$) and
$2J_{||} $ (for $J_\perp/J_{||} =\infty$).
\vskip 0.2cm

The antiferromagnetic spin ladder may be realized in nature by the
antiferromagnet vanadyl pyrophosphate,\cite{JJ,JJGJ} \vopon. This material has
a ladder configuration of $S=1/2$ V$^{+4}$ ions (Fig.2 of Ref.\cite{JJGJ}),
with spacings of 3.19(1) \AA \  between rung ions and 3.864(2) \AA \  between
chain ions, and has a magnetic susceptibility characteristic of an
antiferromagnet with an energy scale (from the susceptibility
maximum) of $\approx 7$ meV. The closely related material \vopoh
has isolated V$^{+4}$ ion pairs at a similar
separation of 3.10 \AA, and is well described magnetically by independent
spin-1/2 Heisenberg pairs with a coupling
(in our conventions) of $J=7.81$ meV.\cite{VOH}

Although the \vopo lattice clearly shows a ladder configuration of V$^{+4}$
ions, they might in principle interact magnetically as a different spin system.
This was the case for Cu(NO$_3$)$_2$$\cdot {5\over 2}$ H$_2$O
(Ref.\cite{cuno}), which was originally considered to be a
possible spin ladder system but was
subsequently found to interact as a dimer chain, described by the Hamiltonian
\begin{eqnarray}
H = \sum_i \bigg\{
J_1 \;{\bf S}_{2i} \cdot {\bf S}_{2i+1}
+
J_2\; {\bf S}_{2i+1} \cdot {\bf S}_{2i+2}
\bigg\} \ ,
\end{eqnarray}

\noindent
with $J_2/J_1\equiv \ad$ ($0\leq\ad\leq 1$).
It was not practical to distinguish between ladder and dimer
chain models of copper nitrate from bulk thermodynamic properties alone,
which were found to be
very similar for the two
systems.\cite{cuno} The issue was finally decided in favor
of the dimer chain model by proton resonance\cite{pmr} and
neutron diffraction\cite{eckert} experiments.

Vanadyl phosphate presents similar ambiguities.
Although its susceptibility
has been measured and is accurately described by
the susceptibility of
a dimer chain\cite{JJGJ} (2), it is widely believed that
the ladder Hamiltonian (1) will lead to a very similar $\chi(T)$ and so
is not excluded by the good agreement with the dimer chain.
Since no theoretical
results have been published for the bulk limit
ladder susceptibility in the
relevant $J_\perp \sim J_{||}$ regime,
comparison of the experimental susceptibility to the ladder model
has not been possible.
In this paper we present new
numerical results for the bulk limit susceptibility of
ladders and dimer
chains, and fit these to the data for
\vopon. As we shall see, these two models do give very
similar results for the susceptibility, and both
give excellent fits to
\vopo
with appropriate parameters. The
ladder $\chi(T)$ is preferred, although the
differences may be less important than the approximations made
in the models.

\newpage

\vskip 0.4cm
\centerline{\bf II. Spin ladder and dimer chain susceptibilities}
\vskip 0.25cm

\noindent
We determined the susceptibility on finite lattices by generating
all energy levels $\{ E_i\} $ and their multiplicities $\{ d_i\} $
in each sector of fixed total $S_z$, using a Householder algorithm.
The susceptibility was
then obtained through its relation to the expected squared magnetization,
summed over energy levels and total $S_z$ sectors;
\begin{eqnarray}
\chi(T) = g^2\mu_B^2\beta\;
{
\sum_{S_z} \sum_i \; S_z^2 \; d_i\; e^{-\beta E_i}
\over
\sum_{S_z} \sum_i \;  d_i\; e^{-\beta E_i} } \ .
\end{eqnarray}
\noindent
This approach has the advantage that explicit eigenvectors are not required.
For the ladder geometry we used (3) to determine $\chi$ on $2\times L$ lattices
with $L=3,4,\dots, 8$ for couplings $J_\perp / J_{||} = \alpha_\ell =
0.5,0.7,0.9,1.0$
and $1.1$ and
for a range of $T/ J_{||}$ values; for reference purposes the $2\times 8$
results are given in Table I. To estimate the bulk limit susceptibility the
$2\times L$ results were fitted to the form $\chi_L = \chi_{\infty}  + a
e^{-bL}$ at each $T$ and coupling, independently for $L=even$ and $L=odd$,
since these approached the bulk limit from opposite directions. This gave
independent estimates of the bulk limit susceptibility, which allowed a test of
the accuracy of our extrapolation in $L$. Our bulk limit estimate was taken to
be the average of the even-$L$ and odd-$L$ extrapolated values. For the dimer
chain we followed a similar procedure for parameter values $J_2/J_1 = \ad =
0.2,0.4,\dots, 1.0$ and $L=4,6,\dots, 16$; Table II gives the $L=16$ results.
For the dimer chain there is no odd/even effect in the long axis (we always
assume an even number of spins so the ground state has no net magnetization),
so we had only a single extrapolation in $L$ for each $T$. This was compensated
by smaller finite size artifacts than on the ladder, because the long axis of
the dimer chain spanned a maximum of 16 rather than 8 spins.
For fitting purposes it is useful to have a parametrization of these results
that allows accurate interpolation in $T$ and interaction strengths. We tested
several forms and found that the six-parameter function
\begin{eqnarray}
\chi(T) =
{c_1\over T}
\; \big[ 1 + (T/c_2)^{c_3}\; \big(e^{c_4/T}-1\big) \; \big]^{-1} \;
\big[ 1 + \big( {c_5/ T} \big)^{c_6} \big]^{-1}
\end{eqnarray}
\noindent
adequately describes both the experimental data\cite{fit} and the theoretical
ladder and dimer chain susceptibilities over a range of parameters
relevant to \vopon. This form also has the advantage that it
incorporates the exponential behavior expected at low temperatures, unlike
other parametrizations used previously for the dimer chain, and at high
temperature it gives the correct Curie form
$g^2\mu_B^2 / 4 k_B T$.
(The overall normalization $c_1$ is identically
equal to $ g^2\mu_B^2/4k_B$
in all cases, but the $V^{4+}$ ion $g$-factor is unknown {\it a
priori} and is determined when $c_1$ is fitted to the data.)

For the two theoretical susceptibilities we fitted our numerical bulk limit
results to (4), with each of the coefficients $c_2,\dots,c_6$ taken to be
quadratic in the ratio of the two coupling constants.
Preliminary fits to the \vopo susceptibility indicated that the values
$\alpha_\ell = 1.0$ and $\alpha_d = 0.7$ were close to optimum, so we
parametrized our bulk limit results in terms of the departure from these
values.
The fitted coefficients $c_2\dots c_6$
for the ladder over the range $0.9\lapx \al \lapx 1.1$ (with $J_{||}=1$)
were
found to be
\begin{mathletters}
\begin{eqnarray}
c_2 = +2.315 - 4.035\dal + 7.050\dal^2    \\
c_3 = +0.403 - 1.025\dal + 1.850\dal^2    \\
c_4 = +0.443 + 0.225\dal + 0.850\dal^2    \\
c_5 = +0.745 - 0.390\dal + 0.200\dal^2    \\
c_6 = +1.628 - 0.110\dal + 0.600\dal^2
\end{eqnarray}
\end{mathletters}
and for the dimer chain (with $J_1=1$)
over the range $0.6\lapx \ad \lapx 0.8$ we found
\begin{mathletters}
\begin{eqnarray}
c_2 = +8.145 - 61.76\dad + 406.3\dad^2    \\
c_3 = +0.562 + 2.840\dad + 7.300\dad^2    \\
c_4 = +0.456 - 0.435\dad + 1.750\dad^2    \\
c_5 = +0.592 + 1.090\dad + 0.400\dad^2    \\
c_6 = +1.663 + 0.160\dad - 1.40\dad^2  \ .
\end{eqnarray}
\end{mathletters}
Due to the presence of large coefficients this parametrization is not
useful far from the parameter ranges cited; if required the coefficients
could be determined directly from the bulk-limit numerical results.

Our numerical
results for the extrapolated bulk limit susceptibility of the ladder and
dimer chain are
shown in Figs.1 and 2 respectively, together with
the interpolating functions defined by (4-6). The interpolating functions
reproduce the bulk limit susceptibility
with a typical accuracy of
a few times $10^{-4}$
over the parameter ranges quoted above.

\vskip 0.4cm
\centerline{\bf III. Comparison with the experimental \vopo susceptibility}
\vskip 0.25cm

In a previous study Johnston {\it et al.}\cite{JJGJ} presented results for the
susceptibility of \vopon, and noted that the susceptibility of a dimer spin
chain gives a very good description of the data for $\ad\equiv J_2/J_1 =0.7$
(fixed from an interpolation of theoretical curves for 0.6 and 0.8 from
Refs.\cite{cuno}), $J_1=11.32$ meV (hence $J_2 = 7.93$ meV) and $g=2.00$. (Note
that in our conventions the $\{ J_n\} $
are twice as large as in Refs.\cite{JJ} and
\cite{JJGJ}.) A similar coupling of $J=7.81$ meV
was determined for
VO(HPO$_4$)$\cdot \frac{1}{2}$ H$_2$O, which consists of isolated V$^{+4}$
dimers,\cite{VOH} and
the $g$ factor of the V$^{+4}$ ion is known to be quite close to 2 from studies
of other vanadium phosphates.\cite{BBC} Of course it is not clear how the
fitted dimer chain parameters relate to \vopo if it proves to be a spin ladder.
\vskip 0.2cm

To confirm these results we fitted our
three-parameter $(J_1,\alpha_d,g)$ dimer chain susceptibility,
described by (4) and (6), to the data of Ref.\cite{JJGJ}, which consists of 606
values of $\chi(T)$ from $T=7.2^o$K to $344.34^o$K. We found the optimum
parameter values to be
$J_1=11.11$ meV,
$\ad=0.722$
and $g=1.99$.
These are
essentially the parameters found by Johnston {\it et al.}, and the minor
differences are presumably due to
the systematic errors in interpolation (perhaps $1\%$ in parameter values).
The fitted dimer chain
susceptibility and the data for \vopo are shown in Fig.3.

We similarly fitted the three-parameter ($J_{||},\al, g$) ladder
susceptibility (4), (5) to the experimental \vopo $\chi(T)$ data over the full
temperature range. The optimum ladder parameters were found to be
\begin{mathletters}
\begin{eqnarray}
J_{||}=7.76\  {\rm meV} \ ,      \\
\al \equiv J_\perp / J_{||}= 1.007 \ ,      \\
g = 2.03\ .
\end{eqnarray}
\end{mathletters}
The proximity of $g$ to 2 provides a plausibility test of the fit, as does the
fitted value of
$J_\perp = 7.82$ meV,
which is almost identical to
the isolated-dimer
$J = 7.81$ meV
found
previously\cite{VOH} in \vopohn.
These
results suggest that \vopo is very close to a uniform ladder ($J_\perp =
J_{||}$), which is presumably accidental because the rungs are bridged by two
oxygens, whereas the chains have only single oxygens between V$^{+4}$ ions. The
fitted ladder $\chi(T)$ is shown in Fig.4, and this model evidently
also gives a very good
description of the experimental data. The goodness of fit,
defined by the residual $f=\sum_i (\chi_{expt.}(T_i) - \chi_{thy.}(T_i))^2$,
slightly favors the ladder model over the dimer chain.
We cannot choose
between the models definitively from
the susceptibility data,
however, because
the variation in \vopo susceptibility
estimated from samples with
different annealing histories (Fig.1b of Ref.(\cite{JJ})) is somewhat
larger than the difference between the predictions of the
ladder and dimer chain models.

Finally, to test how well $J_\perp$ and
$J_{||}$ are determined, we studied the residual $f$ in constrained
two-parameter fits with $g$ and $J_{||}$ variable but $\al = J_\perp/J_{||}$
fixed. As we changed $\al$ from the optimum value 1.007 we found that by
0.90 and 1.12 the residual had increased by a factor of two.
As we increase $\al$ through the range $[0.9,1.1]$ the
fitted value of $J_{||}$ decreases from 8.2 meV to 7.3 meV, which
can be taken as a conservative estimate
of the accuracy to which $J_{||}$ is determined by the susceptibility.
The fitted $g$
factor remains close to 2.03 over this range.
Outside this
range of $\al$ there is a rapid decrease in the quality of fit,
reaching a factor of five
increase in $f$ by $\al = 0.81$ and 1.24.

\vskip 0.4cm
\centerline{\bf IV. Predictions of the ladder and dimer models}
\vskip 0.25cm

Since we have determined ladder parameters for \vopo from our fit to the
susceptibility, we can use the results of Ref.\cite{BDRS} to give
predictions for the gap and spin-wave excitation spectrum. From Fig.2 of that
reference we can see that the gap near $J_\perp / J_{||} =1$ is quite well
determined by the Lanczos and Monte Carlo studies. An approximate linear
interpolation gives
\begin{eqnarray}
{E_{gap} \over  J_{||}}\bigg{|}_{\al \approx 1}
\approx 0.50(1) + 0.65\; (\al -1) \ ,
\end{eqnarray}
\noindent
so for the optimum fitted parameters we predict
\begin{eqnarray}
E_{gap} = 3.9(1) \ {\rm meV} \ .
\end{eqnarray}
\noindent
Over the parameter range $0.9\leq J_\perp / J_{||} \leq 1.1$ discussed
above the predicted gap increases from 3.6 meV to 4.1 meV.\cite{Roger}

As was noted in Fig.3 of Ref.\cite{BDRS}, for $J_\perp = J_{||}$
the minimum energy required to excite a triplet spin wave
on the
ladder as a function of $k$
closely resembles the dispersion relation of a
spin-1/2 chain, except for the presence of excitation gaps.
The lowest excitation is at $k=\pi$, where the gap is
$\omega(\pi)=0.50 J_{||}$, the maximum is shifted to a $k< \pi /2$, and a
secondary minimum is at $k=0$. This dispersion relation is symmetric about
$k=\pi$.

A complication not noted in Ref.\cite{BDRS} is that the lowest-lying triplet
spin-waves with these parameters arise from
two distinct bands.
The ``primary" band, which
contains the lowest gap, is odd under chain interchange $(k_\perp = \pi )$,
and is shown as solid lines
for $J_\perp / J_{||} = 0.5, 1.0$ and 2.0 in Fig.5. For large $J_\perp$ these
are excitations of a single rung, with energy $\omega\approx J_\perp$.
The ``secondary" band (dashed lines in Fig.5) is even under chain interchange,
and for
large $J_\perp$ these states consist of two excited rungs (hence
$\omega'\approx 2J_\perp$
and the even symmetry), with the two $S=1$ excited rungs
coupled to give $S_{tot}=1$. Thus the secondary band may be interpreted
as the excitation of two spin-wave quanta of the primary band.
This interpretation leads us to anticipate several features of the
secondary dispersion relation in the bulk limit,
for example $\omega'(k=0) = 2\, \omega(k=\pi)$, so the band minimum of the
secondary band in \vopo should lie at 7.8(2) meV given our parameters.
One may similarly construct the entire secondary $\omega'(k')$ given
the primary $\omega(k)$ (assuming there are no bound states),
by finding the minimum-energy combination of two
quanta with specified $k'$.

In our representation in Fig.5 we
fitted the function
\begin{eqnarray}
\omega(k) = [
\omega(0)^2\cos^2(k/2)
+
\omega(\pi )^2\sin^2(k/2)
+ c_0^2\sin(k)^2 ]^{1/2}  \ ,
\end{eqnarray}
\noindent
which interpolates between the known analytic
chain and dimer limits,
to the 2x12 lattice data. (Except for the points $\omega'(k=0)$
in the secondary band,
which showed large finite size effects, and which we replaced as argued above
by $2\, \omega(\pi)$.)
For $J_\perp = J_{||}=J$ the fitted constants were found to be
$\omega(0)=1.890 J$,
$\omega(\pi)=0.507 J$ and
$c_0=1.382 J$.
In Fig.6 we show the triplet dispersion relation
which this parametrization predicts for \vopon,
together with a similar result for the secondary band, using the mean value
$J_\perp = J_{||} = 7.79$ meV and the physical lattice spacing. The primary
triplet band extends from 3.9 meV
at $k=0.813 A^{-1}$ to 16 meV at about $0.3 A^{-1}$, and then falls to 15
meV at $k=0$.
The secondary band
extends from 7.9 meV at $k=0$ to a broad plateau at an energy of
about 17-18 meV centered
on $k=0.813 A^{-1}$. Structure function calculations on the 2x12
lattice suggest that the secondary band should appear most clearly near
the $k=\pi$ point ($0.813 A^{-1}$).

For comparison we quote predictions for the triplet spin-wave dispersion
relation in the dimer chain model. Of course the lattice spacing $a$ and the
direction of the continuous momentum variable $k$ are problematical for
\vopo in the dimer model
because there is no obvious dimer chain interaction pathway. Since the dimer
unit cell has length $2a$ the dispersion relation repeats with period $\Delta k
= \pi /a$; this implies that the two different gaps we found for the ladder
at 0 and $\pi / a$ are equal in the dimer chain. Another characteristic
feature of the dimer chain dispersion relation is that it is symmetric about
$\pi / 2a$, due to inversion symmetry. For the parameters $J_1$, $J_2$ and $g$
found in our susceptibility fits the dimer chain model predicts a
somewhat larger gap of $E_{gap}\approx 0.44 J_1$ = 4.9
meV and a bandwidth of $\approx 11$ meV. It is interesting that one can
apparently distinguish between the dimer chain and ladder models by an accurate
measurement of the gap alone, using parameters derived from susceptibility
fits.

\centerline{\bf V. Summary and Conclusions}
\vskip 0.25cm
In this paper we used numerical techniques to study the susceptibility of a
Heisenberg antiferromagnetic spin ladder and a dimerized Heisenberg spin chain.
We used exact numerical diagonalization to generate
all energy eigenvalues and their degeneracies, which were then used
to determine $\chi(T)$ on ladders and dimer
chains of up to 16 spins.
We presented results
for a range of temperatures and interaction ratios
$J_\perp / J_{||}$ (ladder) and $J_2/J_1$ (chain). These were extrapolated to
give bulk limit estimates, which we parametrized using a function with five
parameters. We fitted the bulk limit $\chi(T)$ to the susceptibility data for
\vopon, which is a candidate spin ladder system but is known to be
accurately described by the dimer chain susceptibility. Our best fit to the
dimer chain model accurately reproduces previous parameter values. Our best fit
for the ladder is in slightly better agreement with the data, and indicates
that
\vopo has very similar $J_\perp$ and $J_{||}$ values. With these parameters we
give numerical predictions for the spin-wave excitation gap of \vopo and for
other
properties of the spin-wave dispersion relation.

\vskip 0.5cm
\centerline{\bf VI. Acknowledgements}
\vskip 0.25cm

We are grateful to D.C.Johnston for providing us with the \vopo susceptibility
data shown in Figs.3 and 4. We thank A.J.Berlinsky,
E.Dagotto, R.Eccleston and D.A.Tennant
for additional useful discussions relating to this work.
The hospitality of
the Theory Group of the Rutherford Appleton Laboratory in
the course of this work is gratefully acknowledged.
We also acknowledge allocations of Cray YMP CPU time at the Supercomputer
Computations Research Institute in Tallahassee and at the
National Supercomputer Center in Urbana.
This study was carried out as part of the Quantum
Structure of Matter Project at Oak Ridge National Laboratory, funded by the
USDOE Office of Scientific Computation under contract DE-AC05-840R21400,
managed by Martin Marietta Energy Systems Inc. and contract DE-FG05-87ER40376
with Vanderbilt University.

\newpage

\begin{table}
\caption{$\chi(T)$ versus T/J$_{\parallel}$ for the $2 \times 8$
ladder model.}

\begin{tabular}{|c|l|c|c|c|c|}
    &
 \multicolumn{5}{c|}{$\alpha_{\ell} = J_{\perp} / J_{\parallel}$} \\
\cline{2-6}
  T/J$_{\parallel}$   &   0.5   &  0.7  & 0.9  &   1.0 &   1.1  \\ \hline
 0.05    &   .001034 &   .000434 &   .000095 &   .000036 &   .000012 \\
 0.10    &   .024049 &   .015862 &   .007576 &   .004697 &   .002730 \\
 0.15    &   .052841 &   .040907 &   .026020 &   .019306 &   .013692 \\
 0.20    &   .071958 &   .059399 &   .043488 &   .035436 &   .027946 \\
 0.25    &   .086240 &   .073189 &   .057624 &   .049470 &   .041489 \\
 0.30    &   .098180 &   .084851 &   .069643 &   .061651 &   .053662 \\
 0.35    &   .108058 &   .094958 &   .080161 &   .072384 &   .064535 \\
 0.40    &   .115854 &   .103404 &   .089220 &   .081722 &   .074105 \\
 0.45    &   .121734 &   .110128 &   .096743 &   .089607 &   .082308 \\
 0.50    &   .125983 &   .115240 &   .102749 &   .096035 &   .089124 \\
 0.55    &   .128895 &   .118955 &   .107366 &   .101100 &   .094615 \\
 0.60    &   .130729 &   .121508 &   .110776 &   .104954 &   .098906 \\
 0.65    &   .131696 &   .123113 &   .113171 &   .107773 &   .102151 \\
 0.70    &   .131970 &   .123957 &   .114733 &   .109728 &   .104508 \\
 0.75    &   .131691 &   .124192 &   .115617 &   .110972 &   .106127 \\
 0.80    &   .130975 &   .123942 &   .115954 &   .111638 &   .107136 \\
 0.85    &   .129917 &   .123308 &   .115852 &   .111835 &   .107648 \\
 0.90    &   .128592 &   .122373 &   .115402 &   .111656 &   .107754 \\
 0.95    &   .127064 &   .121205 &   .114674 &   .111175 &   .107534 \\
 1.00    &   .125384 &   .119856 &   .113728 &   .110454 &   .107051 \\
 1.25    &   .115891 &   .111691 &   .107134 &   .104726 &   .102238 \\
 1.50    &   .106267 &   .102995 &   .099495 &   .097661 &   .095775 \\
 1.75    &   .097456 &   .094849 &   .092086 &   .090649 &   .089174 \\
 2.00    &   .089662 &   .087541 &   .085311 &   .084156 &   .082975 \\
 2.25    &   .082839 &   .081084 &   .079250 &   .078303 &   .077337 \\
 2.50    &   .076876 &   .075402 &   .073867 &   .073077 &   .072274 \\
 2.75    &   .071649 &   .070394 &   .069092 &   .068424 &   .067745 \\
 3.00    &   .067046 &   .065965 &   .064848 &   .064276 &   .063695 \\
 3.25    &   .062971 &   .062031 &   .061062 &   .060566 &   .060064 \\
 3.50    &   .059344 &   .058520 &   .057671 &   .057238 &   .056799 \\
 3.75    &   .056099 &   .055370 &   .054621 &   .054239 &   .053853 \\
 4.00    &   .053181 &   .052531 &   .051866 &   .051527 &   .051184 \\
\end{tabular}
\end{table}

\newpage

\mediumtext
\begin{table}
\caption{$\chi(T)$ versus T/J$_1$ for the ${\rm N} = 16$ dimer chain
model.}

\begin{tabular}{|c|c|c|c|c|c|c|}
         &  \multicolumn{6}{c|}{$\alpha_d = J_2 / J_1$}       \\ \cline{2-7}
  T/J$_1$  &  0.2   &  0.4   &   0.6   & 0.7  &  0.8   &  1.0  \\ \hline
 0.05  &  .000000 &  .000001 &  .000032 &  .000207 &  .001380 & .011118  \\
 0.10  &  .000616 &  .001530 &  .006061 &  .013706 &  .031256 & .073202  \\
 0.15  &  .009732 &  .015373 &  .032088 &  .049187 &  .073849 & .105906  \\
 0.20  &  .035487 &  .045476 &  .067867 &  .084726 &  .103142 & .117932  \\
 0.25  &  .071654 &  .081545 &  .100186 &  .111536 &  .121656 & .124279  \\
 0.30  &  .107862 &  .114179 &  .125146 &  .130670 &  .134374 & .129372  \\
 0.35  &  .138004 &  .139715 &  .143241 &  .144277 &  .143674 & .134013  \\
 0.40  &  .160397 &  .158026 &  .155844 &  .153834 &  .150521 & .138121  \\
 0.45  &  .175673 &  .170271 &  .164219 &  .160299 &  .155375 & .141511  \\
 0.50  &  .185222 &  .177821 &  .169378 &  .164351 &  .158550 & .144073  \\
 0.55  &  .190459 &  .181882 &  .172110 &  .166518 &  .160317 & .145789  \\
 0.60  &  .192568 &  .183407 &  .173028 &  .167222 &  .160925 & .146707  \\
 0.65  &  .192465 &  .183123 &  .172604 &  .166806 &  .160603 & .146916  \\
 0.70  &  .190827 &  .181567 &  .171200 &  .165544 &  .159549 & .146517  \\
 0.75  &  .188146 &  .179129 &  .169092 &  .163656 &  .157931 & .145614  \\
 0.80  &  .184775 &  .176099 &  .166488 &  .161314 &  .155889 & .144304  \\
 0.85  &  .180966 &  .172681 &  .163547 &  .158652 &  .153537 & .142674  \\
 0.90  &  .176898 &  .169028 &  .160386 &  .155772 &  .150965 & .140797  \\
 0.95  &  .172699 &  .165246 &  .157093 &  .152755 &  .148245 & .138736  \\
 1.00  &  .168456 &  .161412 &  .153734 &  .149660 &  .145433 & .136543  \\
 1.25  &  .148209 &  .142928 &  .137255 &  .134274 &  .131202 & .124795  \\
 1.50  &  .130981 &  .126962 &  .122690 &  .120461 &  .118172 & .113424  \\
 1.75  &  .116819 &  .113687 &  .110382 &  .108667 &  .106912 & .103283  \\
 2.00  &  .105181 &  .102682 &  .100061 &  .098707 &  .097324 & .094474  \\
 2.25  &  .095528 &  .093493 &  .091370 &  .090276 &  .089161 & .086869  \\
 2.50  &  .087431 &  .085744 &  .083990 &  .083090 &  .082173 & .080293  \\
 2.75  &  .080559 &  .079139 &  .077668 &  .076914 &  .076147 & .074579  \\
 3.00  &  .074663 &  .073452 &  .072201 &  .071561 &  .070912 & .069584  \\
 3.25  &  .069555 &  .068511 &  .067434 &  .066885 &  .066327 & .065190  \\
 3.50  &  .065090 &  .064181 &  .063245 &  .062768 &  .062284 & .061299  \\
 3.75  &  .061157 &  .060357 &  .059537 &  .059119 &  .058696 & .057834  \\
 4.00  &  .057666 &  .056959 &  .056233 &  .055864 &  .055491 & .054731  \\
\end{tabular}
\end{table}


\begin{references}

\bibitem{Haldane} F.D.M.Haldane, Phys. Lett. 93A, 464 (1983);
see also I.Affleck, J. Phys. 1, 3047 (1989).

\bibitem{revtb}
T.Barnes, Int. J. Mod. Phys. C2, 659 (1991).

\bibitem{reved}
E.Dagotto, Int. J. Mod. Phys. B5, 907 (1991).

\bibitem{revem}
E.Manousakis, Rev. Mod. Phys. 63, 1 (1991).

\bibitem{DRS} E.Dagotto, J.Riera and D.Scalapino, Phys. Rev. B45, 5744
(1992); see also
E.Dagotto and A.Moreo,
Phys. Rev. B38, 5087 (1988); Phys. Rev. B44, 5396(E) (1991).


\bibitem{frung}
K.Hida, J. Phys. Soc. Jpn. 60, 1347 (1991);
H.Watanabe,
{\it Numerical Diagonalization Study of an S=1/2 Ladder Model with Open
Boundary Conditions}, University of Tsukuba report;
H.Watanabe, K.Nomura and S.Takada,
J. Phys. Soc. Jpn. 62, 2845 (1993);
H.Watanabe,
{\it Haldane Gap and the Quantum Spin Ladder},
University of Tsukuba Ph.D. thesis (1994).

\bibitem{BDRS} T.Barnes, E.Dagotto, J.Riera and E.S.Swanson,
Phys. Rev. B47, 3196 (1993).

\bibitem{JJ} D.C.Johnston and J.W.Johnson, J. Chem. Soc., Chem. Comm. 1720
(1985); for earlier experimental references see N.Middlemiss,
McMaster University Ph.D. thesis (1978);
Y.Gorbunova and S.A.Linde, Dokl. Akad. Nauk SSSR 245, 584 (1979).

\bibitem{JJGJ} D.C.Johnston, J.W.Johnson, D.P.Goshorn and A.J.Jacobson,
Phys. Rev. B35, 219 (1987).

\bibitem{VOH} J.W.Johnson, D.C.Johnston, A.J.Jacobson and J.F.Brody,
J. Am. Chem. Soc. 106, 8123 (1984).


\bibitem{cuno}
J.C.Bonner, S.A.Friedberg, H.Kobayashi and B.E.Myers,
in {\it Proceedings of the 12th International Conference on Low Temperature
Physics}, ed.E.Kanda (Keigaku, Tokyo 1971), p.691;
J.C.Bonner and H.W.J.Bl\"ote, Phys. Rev. B25, 6959 (1982);
J.C.Bonner, S.A.Friedberg, H.Kobayashi, D.L.Meier and
H.W.J.Bl\"ote, Phys. Rev. B27, 248 (1983).

\bibitem{pmr} K.M.Diederix, J.P.Groen, L.S.J.M.Henkens, T.O.Klassen
and N.J.Poulis, Physica 94B, 9 (1978).

\bibitem{eckert} J.Eckert, D.E.Cox, G.Shirane, S.A.Friedberg
and H.Kobayashi, Phys. Rev. B20, 4596 (1979).

\bibitem{fit} A fit of this form to the experimental susceptibility gave
parameter values of
$c_1 = 0.400$ cm$^3$ ${}^o$K / mole V,
$c_2=201. {}^o$K,
$c_3=0.728$,
$c_4=46.2 {}^o$K,
$c_5=74.1 {}^o$K,
and
$c_6=1.750$. These values were not used
in our fits to the ladder and dimer chain models
and are quoted for reference only.

\bibitem{BBC} D.Ballutaud, E.Bordes and P.Courtine, Mater. Res. Bull. 17,
519 (1982).

\bibitem{Roger} After we obtained our result $E_{gap}\approx 4.$
meV, a gap close to this value was observed in an inelastic neutron scattering
experiment at the ISIS facility, Rutherford Appleton Laboratory,
at the momentum expected in the ladder model,
$k = \pi a^{-1} \approx 0.8$ \AA$^{-1}$. (R.Eccleston
{\it et al.}, in preparation).

\end{references}
\end{document}